# Tuning fulleride electronic structure and molecular ordering via variable layer index

Yayu Wang*, Ryan Yamachika, Andre Wachowiak, Michael Grobis,
and Michael F. Crommie*

*Department of Physics, University of California at Berkeley,
and Materials Sciences Division, Lawrence Berkeley Laboratory,
Berkeley, California 94720-7300*

*e-mail: yywang@berkeley.edu; crommie@berkeley.edu

**$C_{60}$ fullerides are uniquely flexible molecular materials that exhibit a rich variety of behavior[1], including superconductivity and magnetism in bulk compounds[2,3], novel electronic and orientational phases in thin films[4-9], and quantum transport in a single-$C_{60}$ transistor[10]. The complexity of fulleride properties stems from the existence of many competing interactions, such as electron-electron correlations, electron-vibration coupling, and intermolecular hopping. The exact role of each interaction is controversial due to the difficulty of experimentally isolating the effects of a single interaction in the intricate fulleride materials. Here we report a unique level of control of the material properties of $K_xC_{60}$ ultra-thin films through well-controlled atomic layer indexing and accurate doping concentrations. Using STM techniques, we observe a series of electronic and structural phase transitions as the fullerides evolve from two-dimensional monolayers to quasi-three-dimensional multilayers in the early stages of layer-by-layer growth. These results demonstrate the systematic evolution of fulleride electronic structure and molecular ordering with variable $K_xC_{60}$ film layer index, and shed new light on creating novel molecular structures and devices.**

The ability to tune competing interactions in the fullerides arises from advances in our ability to grow well-controlled heterogeneous molecular films. Here we describe measurements on potassium doped $C_{60}$ ($K_xC_{60}$) ultra-thin films having variable thickness from one to three layers (layer index $i = 1$, 2, and 3) for three specific doping concentrations ($x = 3$, 4, and 5). Fig. 1a displays a scanning tunneling microscope (STM) topograph of a representative $K_xC_{60}$ multilayer on Au(111), where the color scale highlights the plateau structure. Narrow slivers of $C_{60}$-free voids containing only K atoms (brown) exist between continuous patches of $K_xC_{60}$. Islands of second (blue) and third layer (red) $K_xC_{60}$ can be seen residing on top of the first $K_xC_{60}$ layer (green). The average layer thickness is ~9.9 Å, greater than the 8 Å spacing found in undoped $C_{60}$ films[11].

We begin by describing our results for a multilayer of the *x* = 3 metallic system. Layer-dependent electronic structure in $K_3C_{60}$ can be seen in Fig. 2a, which shows spatially-averaged *dI/dV* spectra measured at three different layer levels. Within each layer the spectrum is highly uniform with no sign of spatial inhomogeneity such as that found in the surface of bulk fullerides[12]. The first layer *dI/dV* displays a wide peak at the Fermi energy ($E_F$), reflecting the large electronic density of states (DOS) of a metallic LUMO-derived band (LUMO = Lowest Unoccupied Molecular Level). In contrast, the second layer spectrum shows a sharp dip at $E_F$, indicating the emergence of an energy gap that tends to split the band into two halves. A similar gap-like feature persists in the third layer. The width of the gap-like feature (measured between adjacent local maxima) is ~ 0.2 eV, a much larger value than the superconducting gap $2\Delta_{sc}$ ~ 6 meV found in bulk $K_3C_{60}$[13].

The spatial arrangement of $C_{60}$ molecules also changes dramatically with layer index. The first layer of $K_3C_{60}$ (Fig. 2c) exhibits a complex $\sqrt{3} \times \sqrt{3}$ superstructure of bright molecules having different orientation from their dimmed nearest neighbors[8]. In the second layer (Fig. 2d), however, $C_{60}$ molecules form a very simple hexagonal lattice (lattice constant *a* ~10.5 Å) with long-range orientational ordering. The tri-star-like topography of each molecule suggests that $C_{60}$ in the second layer is oriented with a hexagon pointing up[14]. The third layer topograph is the same as the second layer.

The insulating $x = 4$ multilayer system displays a similar trend. Fig. 3a shows $dI/dV$ spectra measured on a $K_4C_{60}$ plateau structure where the number of layers is varied from $i = 1$ to 3. First layer spectra ($i = 1$) exhibit an insulating energy gap $\Delta \sim 0.2$ eV that is induced by molecular Jahn-Teller (JT) distortion[8]. As the layer index increases from $i = 1$ to 3, the energy gap opens continuously (by layer 3 the gap has well-defined edges and a flat bottom). The gap amplitudes observed here are estimated to be $\Delta \sim 0.6$ eV and 0.8 eV for layer 2 and 3 respectively. As seen in the metallic $x = 3$ system, the geometric structure of the insulating $x = 4$ system simplifies as layer index is increased. Complex "cross-phase" orientational ordering observed in the $K_4C_{60}$ first layer[8] (Fig. 3c) evolves to a much simpler hexagonal lattice (Fig. 3d) for higher layers. Layers 2 and 3 for the $x = 4$ film display featureless $C_{60}$ molecules with little discernable orientational ordering.

As doping is increased to $x = 5$, $K_xC_{60}$ multilayers show re-entrant metallicity for layer 1 and an evolution to insulating behavior by layer 3. At this doping the upper JT-band is only partially filled and the first layer spectrum (Fig. 4a) exhibits a finite (although suppressed) electronic DOS at $E_F$. Suppression of the DOS at $E_F$ deepens significantly in the second layer, and broadens to a pronounced energy gap by the third layer. As with the $x = 3$ and $x = 4$ doping levels, complex structural ordering in the first layer of $K_5C_{60}$ (i.e., the highly ordered 2×2 superstructure seen in Fig. 4c) evolves into a much simpler hexagonal ordering in higher layers (Fig. 4d).

Taken collectively, these results clearly show that increasing film thickness suppresses metallicity and enhances the insulating tendency of $K_xC_{60}$ thin films on Au(111). We propose that this trend arises from electron correlations due to intramolecular Coulomb repulsion, characterized by the Hubbard $U$. In fullerides, $U$ is on the same order as the narrow bandwidth $W^1$. The fullerides thus exist at the verge of a Mott-insulator phase transition and small perturbations to the strength of $U$ may significantly alter their electronic ground state[15-19]. The electronic phase transitions observed here can be shown to result from changes in $U$ wrought by changes in the local screening environment.

We now examine quantitatively the effect of screening on electron correlations in molecular films. For an isolated single $C_{60}$ molecule, the bare Hubbard $U$ (denoted as $U_0$) has been estimated to be approximately 3.0 eV[20-23]. When $C_{60}$ molecules form a solid, the value of $U$ can be greatly reduced by screening. In our $K_xC_{60}$ multilayers screening originates from three distinct sources: the metal substrate, surrounding polarizable molecules, and itinerant charge carriers.

The first two screening mechanisms can be treated in a straight-forward manner. As discussed by Hesper *et al.*[24], the screening of $U$ from a nearby metal can be modeled using an image charge potential as

$$\delta U_S = e^2/(4\pi\varepsilon_0 2d), \qquad (1)$$

where $d$ is the distance from the center of a molecule to the metal substrate (Fig. 1b). The screening from nearby polarized molecules can be expressed as

$$\delta U_P = z\,\alpha\,e^2/(4\pi\varepsilon_0 R^4), \qquad (2)$$

where $z$ is the number of nearest neighbors (*NN*), $\alpha$ is the molecular polarizability, and $R$ is the inter-molecular distance. The third mechanism, renormalization of $U$ by itinerant electrons ($\delta U_e$), depends sensitively on the doping concentrations of the fullerides and is a more complex term[21, 22, 25, 26]. The total reduction of $U$ due to molecular environment is the sum of the three terms:

$$\delta U = \delta U_S + \delta U_P + \delta U_e, \qquad (3)$$

and $U_{eff} = U_0 - \delta U$ is the final effective Hubbard $U$. These relations show how the variable layer structure of our $K_xC_{60}$ ultra-thin films can provide a systematic technique to control electron correlations by changing both the distance to the metal substrate ($d$) and the number of nearest neighbors ($z$).

This is seen most clearly in the insulating $K_4C_{60}$ multilayer, where layer dependence of the gap is accounted for by Eqns. (1)-(2) due to the absence of itinerant charge carriers ($\delta U_e = 0$). As the layer index increases from $i = 1$ to 3, $\delta U_S$ (substrate screening) decreases from 1.5 eV to 0.3 eV according to Eq. (1). In contrast, $z$ increases from 6 in the first layer to 9 for both the second and third layer terraces, causing $\delta U_P$ to increase from 0.6 eV to 0.9 eV via Eq. (2) (here we use $\alpha = 90$ Å$^3$, the value for undoped $C_{60}$)[27]. The total screening (Eq. (3)) is thus $\delta U = 2.1$, 1.4, and 1.2 eV for $i = 1$, 2, and 3, respectively, leading to an increase in $U_{eff}$ from 0.9 eV to 1.8 eV as $i$ increases from 1 to 3. The relative increase of the experimentally observed gap by 0.4 eV (from $i = 1$ to 2) and 0.2 eV (from $i = 2$ to 3) is in fairly good agreement with the calculated relative increase of $U_{eff}$ by 0.7 eV (from $i = 1$ to 2) and 0.2 eV (from $i = 2$ to 3). This agreement becomes much better if we use a larger $\alpha$ value for $K_4C_{60}$ (a reasonable assumption) and near perfect agreement is reached using an $\alpha$ value equal to twice the undoped $C_{60}$ value (unfortunately no independent measure of $\alpha$ has been reported in this regime). Stronger electron correlation in higher layers is thus responsible for the observed layer-dependent increase of gap in $K_4C_{60}$.

The screening behavior in metallic $K_3C_{60}$ multilayers is quite different from the insulating $K_4C_{60}$ case above, since the itinerant electron screening term ($\delta U_e$) becomes important in $K_3C_{60}$. Determining the renormalized $U_{eff}$ for $K_3C_{60}$ is still quite controversial[1, 21, 25]. One way of modeling the screening in this case is to consider a metallic cavity surrounding each $C_{60}$ molecule[1, 26]. This generates a combined screening compensation ($\delta U_e + \delta U_P$) of 2.7±0.5 eV, but does not include substrate effects. When substrate screening ($\delta U_S$) is included using Eq. (1), the effective Coulomb interaction becomes $U_{eff} = U_0 - \delta U$ = 0.2, 0.5 and 0.7 eV for $i$ = 1, 2, and 3. For $i$ = 1 $U_{eff}$ is too small compared to the bandwidth $W$ to open a gap, while in layer 2 and 3 we expect a small Mott-Hubbard gap to begin opening as $U_{eff}$ becomes comparable to $W$ (Fig. 2b). This explains the gap-like structure observed in the $K_3C_{60}$ data of Fig. 2a.

$K_5C_{60}$ is intermediate to the extremes of $K_4C_{60}$ (insulating case) and $K_3C_{60}$ (metallic case). Here itinerant electron screening is greater than for the insulating $K_4C_{60}$, but less effective than the more metallic $K_3C_{60}$. We thus expect the layer dependent $U_{eff}$ for $K_5C_{60}$ to lie between the two extremes, reducing metallicity in the first layer and destroying it by the third. This is indeed the case, as we see a relatively small suppression of electronic DOS in layer 1 (Fig. 4a), but by layer 3 a very well-defined gap has emerged (this has only 60% of the gap magnitude observed for the $i$ = 3 layer in $K_4C_{60}$).

The layer index-dependent change in structural properties seen for all three doping levels as the index increases above $i$ = 1 is less a direct result of electronic screening and more likely dominated by intermolecular electron hopping via the overlap of molecular orbitals[7, 28]. The complex structures found in layer 1 of $K_xC_{60}$, $x$ = 3 to 5 (Figs. 2c, 3c, and 4c) are characteristic of geometric frustration of molecular orientational ordering in a two-dimensional lattice[29]. As the layer index is increased, additional interaction with adjacent $C_{60}$ molecules in the lower layers leads to quasi-3D-like intermolecular interactions. This creates a more isotropic local molecular environment and puts more spatial constraints on molecular orientation, limiting the possibility of exotic molecular ordering. Therefore, the much simpler and more homogeneous spatial structures found in the higher layers of $K_xC_{60}$ (Figs. 2d, 3d, and 4d) can be seen as a natural consequence of dimensional crossover from the 2D limit to the quasi-3D bulk regime.

Using accurately fabricated $K_xC_{60}$ ultra-thin films, we demonstrate how electron correlation strength, a key factor in determining the material properties of fullerides, can be experimentally controlled by varying proximity to a metal substrate, the number of nearest neighbours, and intrinsic doping levels. These results support the notion of tuning molecular electronics via variable layer structure and distance to metal contacts, and open new routes towards engineering novel molecular devices and creating new electronic phases in strongly correlated molecular materials.

## Methods

Our experiments were conducted in a homebuilt ultrahigh vacuum (UHV) cryogenic STM with a PtIr tip. $C_{60}$ thin films with desired thickness were made by evaporating $C_{60}$ molecules onto a clean Au(111) surface from a Knudsen cell evaporator. Appropriate amounts of K atoms were dosed onto the films from a SAES getter. Both K and $C_{60}$ evaporators were pre-calibrated by directly counting the number of atoms/molecules in STM images. The $K_xC_{60}$ thin films were annealed at 200 ºC for 20 minutes before being cooled to 7 K for STM experiments. Progressive doping was obtained by adding more K atoms onto the existing film followed by re-annealing. Lower annealing temperature of

~140 ºC was used in the highly-doped samples ($x > 4$) to avoid K loss. STM topography was carried out in a constant current mode. *dI/dV* spectra were measured through lock-in detection of the ac tunneling current driven by a 450 Hz, 1-10 mV (rms) signal added to the junction bias (defined as the sample potential referenced to the tip) under open-loop conditions.

## Acknowledgements

This work was supported in part by NSF Grant EIA-0205641 and by the Director, Office of Energy Research, Office of Basic Energy Science, Division of Material Sciences and Engineering, U.S. Department of Energy under contract No. DE-AC03-76SF0098. Y. W. acknowledges a research fellowship from the Miller institute for Basic Research in Science.

Figure Captions:

**Figure 1 Structure of a representative $K_xC_{60}$ multilayer thin film on Au(111). a,** Constant current STM topograph of a $K_xC_{60}$ ($x = 4$ here) multilayer thin film on Au(111) taken with sample bias $V = 1$ volt and tunneling current $I = 5$ pA. The plateau structure has four distinct regions: the $C_{60}$-free voids (brown), the first layer (green), second layer (blue), and third layer (red) of $K_xC_{60}$. **b,** Schematic side-view of the multilayer structure. The distances between the centers of each layer to the Au surface are labeled as $d_1$, $d_2$, and $d_3$.

**Figure 2 Electronic and structural properties of a $K_3C_{60}$ multilayer. a,** Spatially-averaged $dI/dV$ spectra measured at three different layer heights of $K_3C_{60}$. The first layer spectrum has a peak at $E_F$, while the second and third layer spectra show a sharp dip with width ~0.2 eV. **b,** Schematic diagram showing the effect of $U$ on the electronic structure of $K_3C_{60}$. The small $U_{eff}$ in first layer (left) has negligible effect on the band structure, but the larger $U_{eff}$ in higher layers (right) causes a dip around $E_F$. **c,** $K_3C_{60}$ first layer image (taken with $V = 0.6$ volt and $I = 20$ pA) shows a $\sqrt{3}\times\sqrt{3}$ superstructure. Each red circle represents one $C_{60}$ molecule. **d,** Topograph of the second layer ($V = 1$ volt and $I = 5$ pA) shows a simple hexagonal lattice.

**Figure 3 Electronic and structural properties of a $K_4C_{60}$ multilayer. a,** *dI/dV* spectra of the JT-insulating $K_4C_{60}$ measured at three different layer heights. The observed gap $\Delta$ increases from 0.2 eV in layer 1 to 0.6 eV and 0.8 eV in layer 2 and 3 due to the enhancement of $U_{eff}$. **b,** Schematic band structure of $K_4C_{60}$. Larger $U_{eff}$ in layer 3 (right) leads to a much wider gap than that in layer 1 (left). **c,** Topograph of $K_4C_{60}$ layer 1 ($V = -0.2$ volt and $I = 5$ pA) shows a "cross"-like orientational ordering. **d,** $K_4C_{60}$ layer 2 displays a simple hexagonal lattice structure with little orientational ordering (image taken with $V = -1$ volt and $I = 10$ pA).

**Figure 4 Electronic and structural properties of a $K_5C_{60}$ multilayer. a,** *dI/dV* spectra of the reentrant metal $K_5C_{60}$ measured at three different layer heights. The electronic structure evolves from metallic in layer 1 to insulating in layer 3. **b,** Schematic electronic structure of $K_5C_{60}$. In layer 1 (left) $U_{eff}$ causes a small suppression of DOS near $E_F$, but in layer 3 (right) the enhanced $U_{eff}$ induces an energy gap around $E_F$. **c,** Topography of layer 1 ($V = -0.2$ volt and $I = 5$ pA) shows complex orientational ordering with a 2×2 superstructure. **d,** $K_5C_{60}$ layer 2 shows simpler hexagonal lattice structure (image taken with $V = -0.2$ volt and $I = 5$ pA).

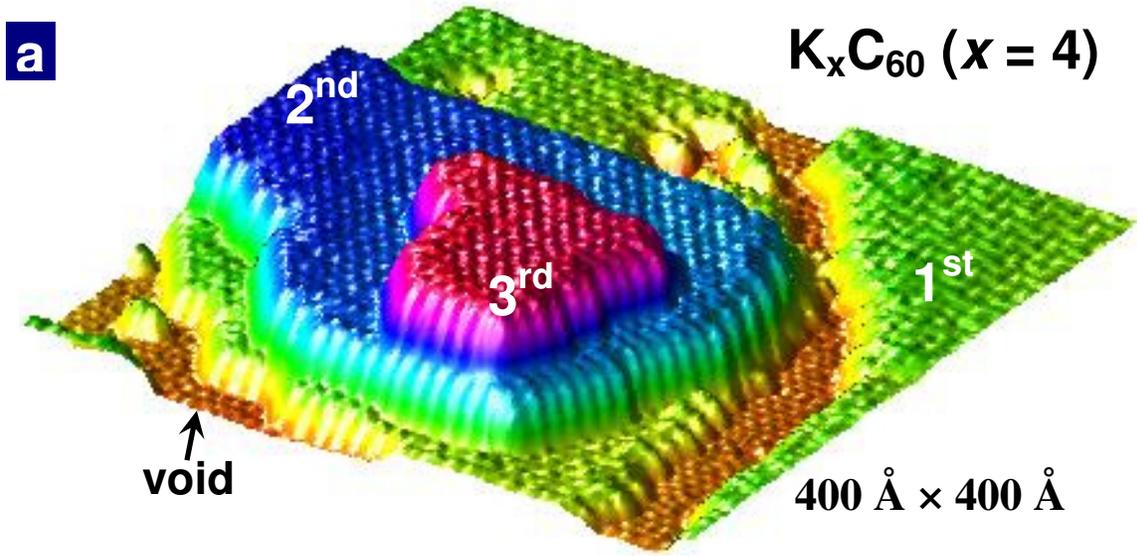
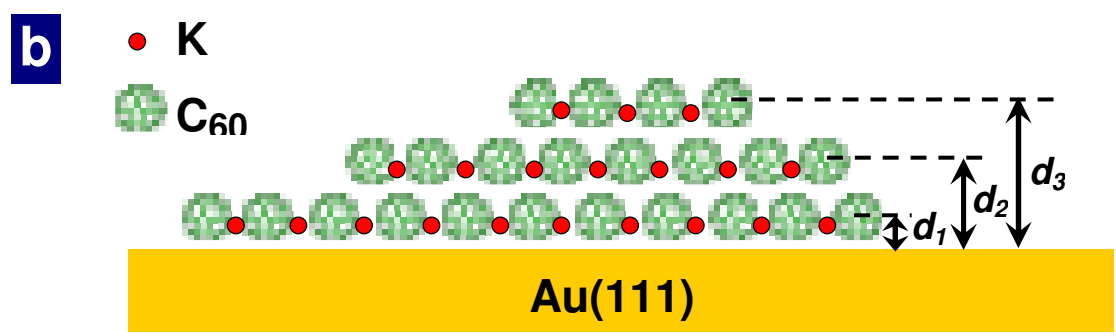

Figure 1

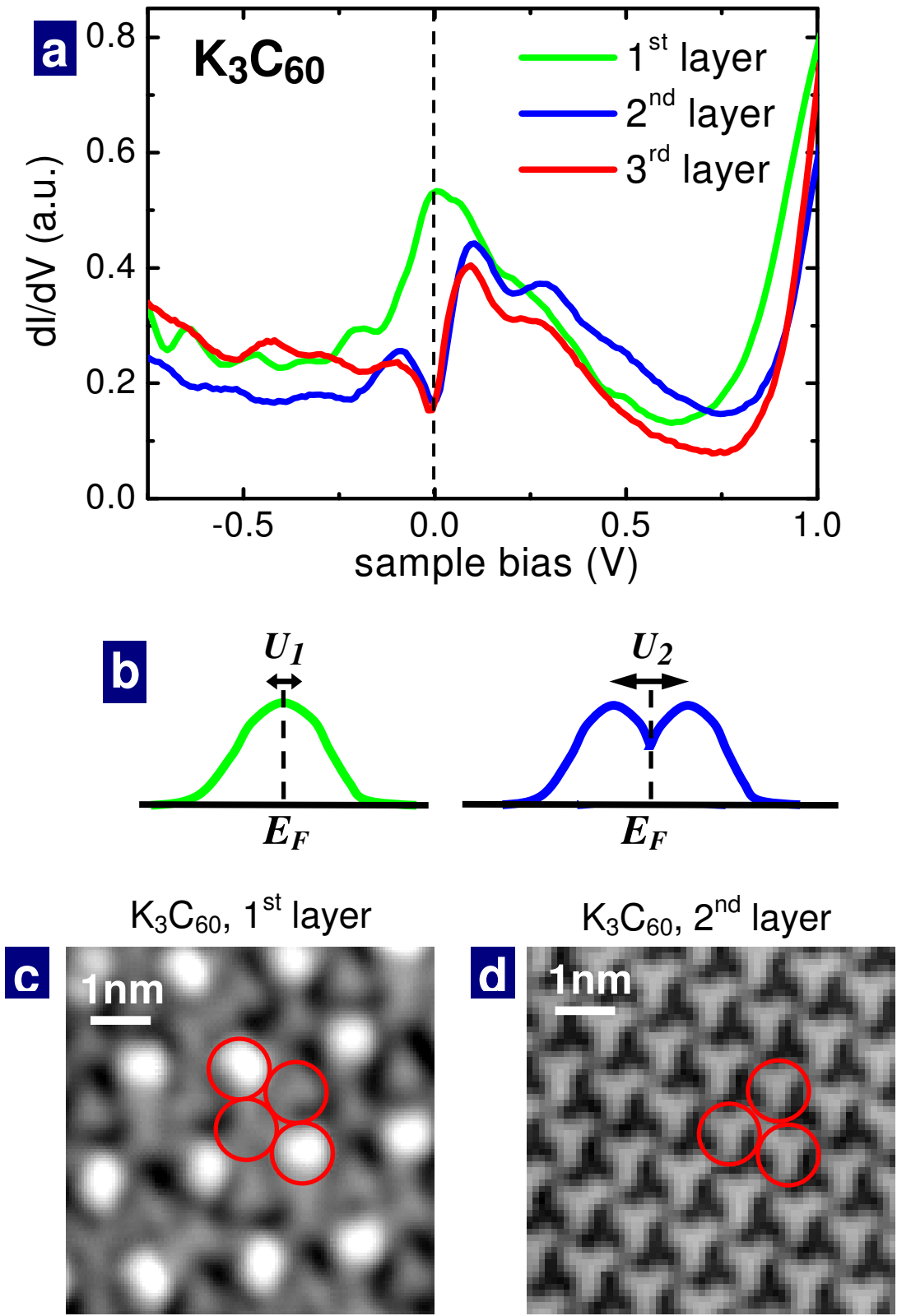

Figure 2

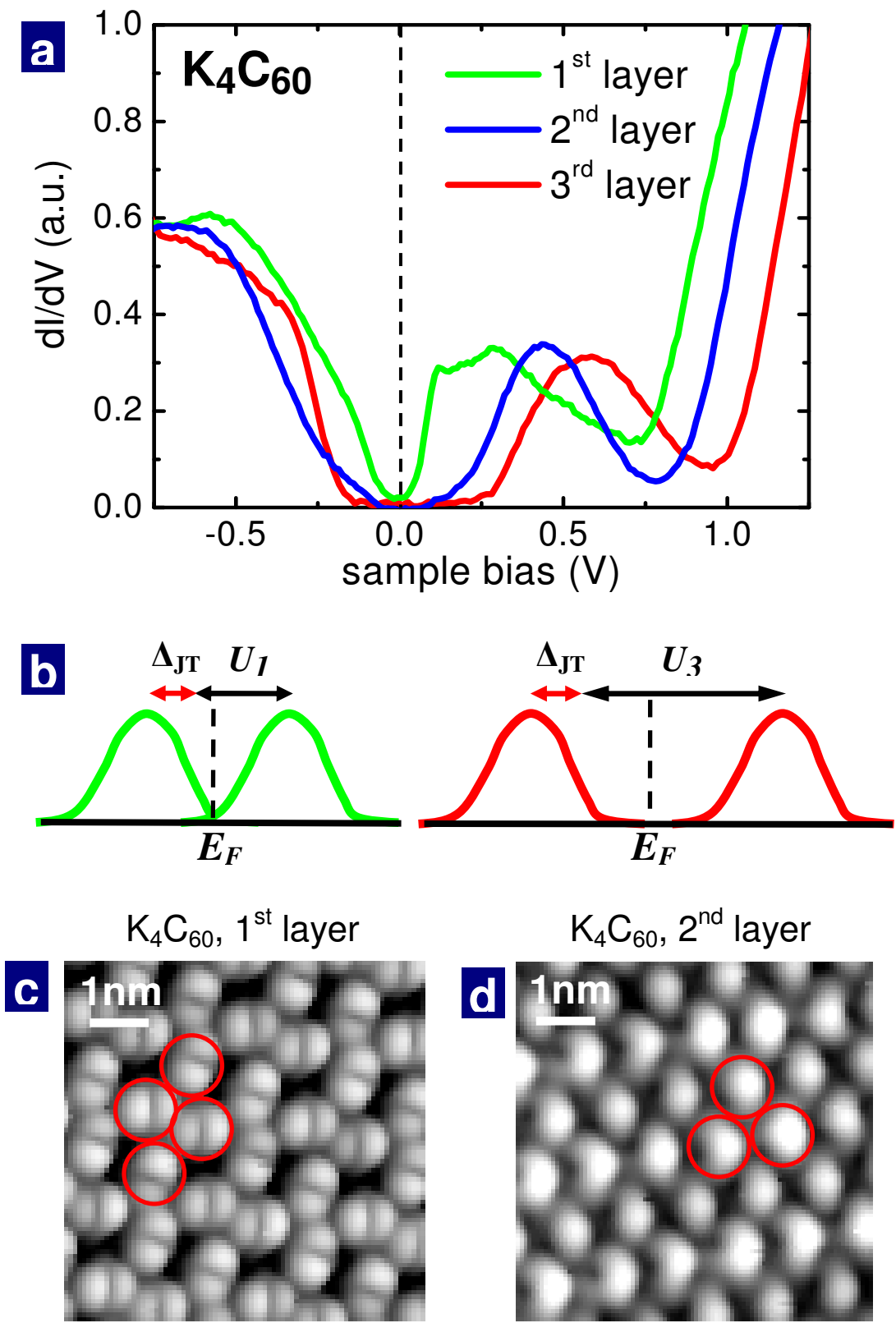

Figure 3

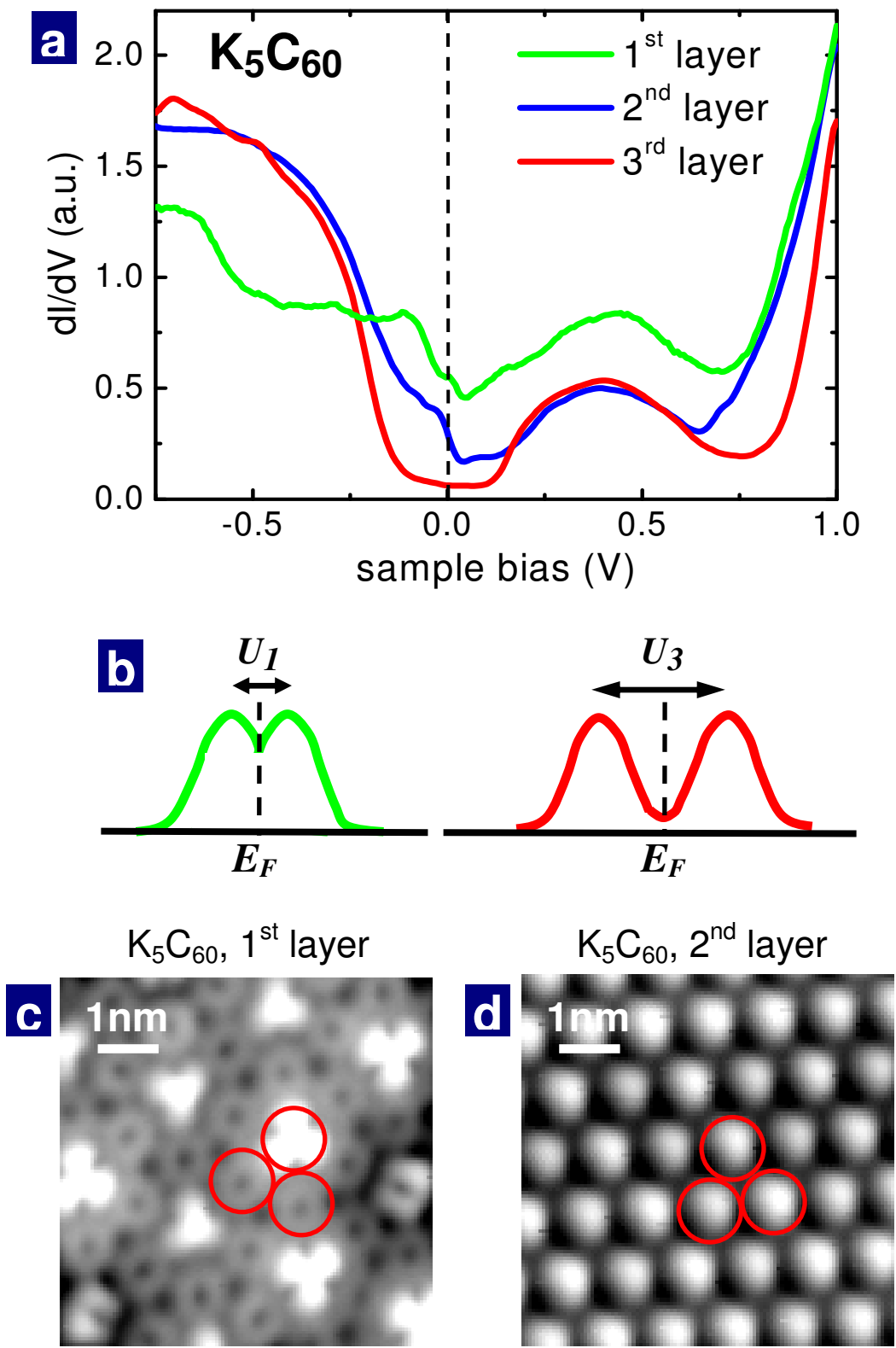

Figure 4